\documentclass[12pt]{article}
\usepackage{amssymb,amsmath,amsthm,amsfonts,amscd}
 \usepackage{graphicx}
\newcommand\be{\begin{equation}}
\newcommand\ee{\end{equation}}
\newcommand\bea{\begin{eqnarray}}
\newcommand\eea{\end{eqnarray}}

\newcommand{\fatalpha}{{\bf \alpha \kern -0.44em \alpha}}
\newcommand{\fatsigma}{{\bf \sigma \kern -0.54em \sigma}}
\newcommand{\tpchi}{{\bf D \kern -0.35em D}}
\newcommand{\llambda}{{\bf \lambda \kern -0.45em \lambda}}

\renewcommand{\theequation}{\arabic{equation}}
\renewcommand{\theequation}{\thesection.\arabic{equation}}
\bibliography{plain}
\title{\bf Quantum Discord for Generalized Bloch Sphere States}\vspace{20mm}
\author{ M. A. Jafarizadeh $^{a,b,c}$
 \thanks{E-mail:jafarizadeh@tabrizu.ac.ir}  ,
  N. Karimi  $^{a,d}$
 \thanks{E-mail:na\_karimi@yahoo.com} and
 H. Zahir  $^{a,b}$
 \thanks{E-mail:choas\_1010@yahoo.com}
 \\ $^a${\small Department of Theoretical Physics and Astrophysics,
Tabriz University, Tabriz 51664, Iran.} \\ $^b${\small Institute
for Studies in Theoretical Physics and Mathematics, Tehran
19395-1795, Iran.} \\ $^c${\small Research Institute for
Fundamental Sciences, Tabriz 51664, Iran. }\\ $^d${\small Farhangian University Allameh amini, Tabriz, Iran.}} \pagebreak

\vspace{20mm}
\begin{document}
\maketitle \vspace{15mm}
\newpage
\begin{abstract}
In this study for particular states of bipartite quantum system in $2^{n}\times 2^{m}$ dimensional Hilbert space state , similar to $m$ or $n$-qubit density matrices represented in Bloch sphere we call them generalized Bloch sphere states(GBSS), we give an efficient optimization procedure so that analytic evaluation of quantum discord can be performed. Using this optimization procedure, we find an exact analytical formula for the optimum positive operator valued measure (POVM) that maximize the measure of the classical correlation for these states. The presented  optimization procedure also is used to show that for any concave entropy function the same POVMs are sufficient for quantum discord of mentioned states.
 Furthermore, We show that such optimization procedure can be used to calculate the geometric measure of quantum discord (GMQD) and then an explicit formula for GMQD is given. Finally, a complete geometric view is presented for quantum discord of GBSS.


{\bf PACs Index: 03.67.-a, 03.65.Ta, 03.65.Ud }
\end{abstract}
\pagebreak

\vspace{7cm}
\section{Introduction}
Quantum correlation and its characterization is an important and one of the most popular re-
search topics but is challenging in quantum information theory. Recently, it was found that quantum correlation  is a key resource in broadcasting of
quantum states \cite{m,h}, quantum state merging \cite{v,d}, assisted
optimal state discrimination \cite{ll}, remote quantum state preparation \cite{ddd}, and so on.

Recently, much effort has been devoted to studying
various measures of non-classical correlation\cite{mp,as,gl,mg,zz,ss,tt,jj}.
As a typical kind of quantum correlation, quantum entanglement \cite{hr} has been well understood in many aspects and widely applied to quantum communication and quantum computation\cite{nil}. It is quite well established that entanglement is essential for certain kinds of quantum-information tasks like quantum cryptography and super-dense coding. However, besides the widely studied feature of entanglement, quantum theory exhibits also
another form of non-classical correlations which is quantified
by the quantum discord(QD) . In an other word, entanglement
only represents a special kind, but not all, of the
quantum correlation.

 Quantum discord captures the nonclassical correlations, including
but not limited to entanglement.
The quantum discord as a measure of quantum correlations, initially introduced
by Ollivier and Zurek \cite{Olliver} and by Henderson and Vedral \cite{vedreal1}, is a measure of the discrepancy between quantum versions of two classically equivalent expressions for mutual information.
Many remarkable applications of QD have  been proposed
such as the characterization of quantum phase transitions \cite{pr} and the dynamics of quantum systems under
decoherence \cite{jm}. Besides, quantum discord could be used to improve the efficiency of the quantum Carnot engine \cite{rd1} and to better
understand the quantum phase transition and the process
of Grover search \cite{rd2,rd3}.
Because evaluation of quantum discord involves optimization procedure, it is difficult
to calculate and it was analytically computed only for some special cases \cite{rd4,rd5,rd6,rd7,rd8}.

To avoid this difficulty and obtain an analytic analysis, Daki$\acute{c}$ et al.\cite{gmm} introduced the GMQD  which defined as the nearest distance between the given state
and the set of zero-discord states. Luo and Fu \cite{lf} introduced another equivalent form for GMQD .

In this paper, we devote to investigate the quantum discord for GBSS in a bipartite system. Our results show that despite of functionality , which is chosen in definition of the entropy for  the quantum discord , the same POVM satisfies  the optimization procedure .
 Also our  analytical calculation of the GMQD, denote that the optimization procedure does not change and  is the same as we have for original quantum discord.\\
The organization of the paper is as follows.
 First, a brief review of the concept of the quantum discord is given. Then, we explain the quantum discord for GBSS states, leading to the obtaining the optimum POVMs and analytical solution of quantum discord.
 Next, an explicit formula for GMQD is presented. Finally, we present  a
geometric viewpoint, from which GBSS quantum discord can be described
clearly. The paper ends with a brief conclusion and one appendix.

\section{The Concept of The Quantum Discord}

Quantum discord \cite{Olliver,vedreal1} is measured by the difference between the mutual information and the maximal conditional mutual information obtained by local measurement. "Right" Quantum discord of a bipartite state $\rho_{AB}$ in a Hilbert space $H^{A}\otimes H^{B}$ is given by\cite{Olliver,vedreal1}
\begin{align} \label{d1}
D_{B}(\rho^{AB})&=I(\rho^{AB})-C_{B}(\rho^{AB}),
  \end{align}
 where
\begin{align} \label{d2}
I(\rho^{AB})&= S(\rho^{A})+S(\rho^{B})-S(\rho^{AB}).
\end{align}

\
Here, $I(\rho^{AB})$ represents the quantum mutual information (the total amount of correlations) of the two-subsystem
state $\rho^{AB}$. $ \rho^{A(B)} = Tr_{B(A)}(\rho^{AB})$ is the reduced density matrix for the subsystem $A(B)$. $S(\rho) = -Tr(\rho log_{2} \rho)$ is the
von Neumann entropy of the system in the state $\rho$.
The other quantity, $C_{B}(\rho^{AB})$, is interpreted as a measure
of the classical correlation of the two subsystems $AB$ in the state $\rho^{AB}$ and it is defined as the maximal information that one can obtain, for example, about $B$ by performing the complete measurement $E_{k}^{B}$ on $H^{B}$,

\begin{align}
C_{B}(\rho^{AB})&=\max_{\lbrace E_{k}^{B}\rbrace}\left[S(\rho^{A})-\Sigma_{k}p_{A|k}S(\rho_{A|k})\right],
\end{align}
where $ \rho_{A|k}=\frac{1}{p_{A|k}}Tr_{B}(I_{A}\otimes E_{k}^{B}\rho^{AB}) $ is the postmeasurement state of $A$ after obtaining the outcome $k$ on $B$ with
the probability $p_{A|k}=Tr[I_{A}\otimes E_{k}^{B}\rho^{AB}]$; maximum is taken over all the von Neumann measurement sets $E_{k}^{B}$ on system B. \\
Similarly, "left" quantum discord is given by

\begin{align} \label{d1}
D_{A}(\rho^{AB})&=I(\rho^{AB})-C_{A}(\rho^{AB}),
  \end{align}
  \begin{align}\label{lef}
C_{A}(\rho^{AB})&=\max_{\lbrace E_{k}^{A}\rbrace}\left[S(\rho^{B})-\Sigma_{k}p_{B|k}S(\rho_{B|k})\right].
\end{align}
where $ \rho_{B|k}=\frac{1}{p_{B|k}}Tr_{A}(E_{k}^{A}\otimes I_{B}\rho^{AB}) $ is the postmeasurement state of $B$ after obtaining the outcome $k$ on $A$ with
the probability $p_{B|k}=Tr[ E_{k}^{A}\otimes I_{B} \rho^{AB}]$.
  Note that the difference between those two discords is that
the measurement is performed on party A or on party B,
respectively. It is thus expected that this definition of quantum
discord is not symmetric with respect to A and B.

  A set of operators $\{E_{k}^{A(B)}\}$ is named
POVM if and only if the following two conditions are met: $(1)$ each operator $E_{k}^{A(B)}$ is positive
positive $\Leftrightarrow \langle\psi|E_{k}^{A(B)}|\psi\rangle \geq 0$, $ \forall|\psi\rangle$	 and $(2)$ the completeness relation
is satisfied, i.e.,
\begin {equation}\label{EFCS1}
\sum_{k}E_{k}^{A(B)}=1.
\end {equation}
The elements of $\{E_{k}^{A(B)}\}$ are called effects or POVM elements.
On its own, a given POVM $\{E_{k}^{A(B)}\}$ is enough to give complete
knowledge about the probabilities of all possible outcomes;
measurement statistics is the only item of interest.

\section{Analytical solution for GBSS}\label{1}
\
Let $\gamma_{\mu}, \mu = 1, ..., d,$ be d Dirac $\gamma$ matrices satisfying the anticommuting relations:
\begin{equation}\label{2111}
\gamma_{\mu}\gamma_{\nu}+\gamma_{\nu}\gamma_{\mu}=2\delta_{\mu\nu}I.\quad \mu,\nu=1,...,d.
 \end{equation}
It follows from relation (\ref{2111}) that the $\gamma$ matrices $\gamma_{\mu}$ generate an algebra which, as a vector
space, has a dimension $2^{d}$ (For a brief review about Dirac matrices and an explicit construction of $\gamma$, we refer the
reader to \cite{xx} or see the Appendix).
 We consider hermitian matrices $\lambda_{i}, i = 1, 2, ..., 2^{d}$ as all possible multiplications of $\gamma_{\mu}, \mu = 1, ..., d$
up to multiplicative factors $±1,\pm i$.\\
 Let a basis for the Lie algebra of $SU(N)$ be given by $\{\lambda_{i}\}_{i=1}^{N^{2}}$.
 We will use the following normalization condition for the elements of the Lie algebra of $SU(N)$
 \begin{align}
                 Tr(\lambda_{i}\lambda_{j})=2\delta_{ij}.
\end{align}
We will also choose the following relations for commutation and anticommutation relations:
$$
[\lambda_{i},\lambda_{j}]=2if_{ijk}\lambda_{k},
 $$
\begin{align}\label{d}
                 \{\lambda_{i},\lambda_{j}\}= \frac{4}{N}\delta_{ij}I+2d_{ijk}\lambda_{k},
\end{align}
where the $f_{ijk}$ are the structure constants and the $d_{ijk}$ are the components of the totally symmetric "d-tensor." These two equations may be combined more succinctly as
\begin{align}
                 \lambda_{i}\lambda_{j}= \frac{2}{N}\delta_{ij}I+if_{ijk}\lambda_{k}+d_{ijk}\lambda_{k}.
\end{align}
Using these conventions, we may express the POVM elements as
$$
E_{e^{k}}^{B}=\frac{1}{M}(I_{M\times M }+\sqrt{\frac{M(M-1)}{2}}e^{k}.\lambda^{B} )= \frac{1}{M}(I_{M\times M}+\sqrt{\frac{M(M-1)}{2}}
   \Sigma_{i=1}^{M^{2}-1} e_{i}^{k}\lambda_{i}^{B} ),
$$
\begin{align}\label{povm}
   E_{e^{k}}^{A}=\frac{1}{N}(I_{N\times N}+\sqrt{\frac{N(N-1)}{2}}e^{k}.\lambda^{A} )= \frac{1}{N}(I_{N\times N}+\sqrt{\frac{N(N-1)}{2}}
   \Sigma_{i=1}^{N^{2}-1} e_{i}^{k}\lambda_{i}^{A} ).
 \end{align}
 This representation is called a coherence vector representation with $e^{k}$ the coherence vector. The constant is a
convenient one such that for pure states
\begin{align}
  e^{k}.e^{k}=1 , e^{k}\star e^{k}=e^{k},
 \end{align}
 where the "star" product is defined by
\begin{align}
  (a\star b)_{l}=\sqrt{\frac{N(N-1)}{2}}\frac{1}{N-2}d_{ijl}a_{i}b_{j}.
 \end{align}

We consider particular states acting on a bipartite system $H^{A}\otimes H^{B}$ with $dim(H^{A})=N = 2^{n}$ and $dim(H^{B})=M = 2^{m}$ which possess
properties similar to $m$ or $n$-qubit density matrices represented in Bloch sphere, and so we call them
generalized Bloch sphere states.\\
 In the case of even dimension d, we denote $\gamma_{s}=i^{-\frac{d}{2}}\gamma_{1}\gamma_{2}...\gamma_{d}$ by $\gamma_{d+1}$, then the matrices $\vec{w}=\{\gamma_{1}, \gamma_{2},..., \gamma_{d}, \gamma_{d+1}\}=\{\gamma_{1}, \gamma_{2},..., \gamma_{2n}, \gamma_{2n+1}=i^{-n}\gamma_{1}\gamma_{2}...\gamma_{2n}\}$ form a maximally anticommuting set in the algebra of $\gamma$ matrices (in the case of odd d, the set of matrices $\gamma_{i}$, i = 1, ..., d is maximally anticommuting set). By choosing maximally anticommuting sets $\vec{w}$ the decomposition of density matrices of A into a
Bloch vector has, in general, the following form:
\begin{equation}
\rho=\frac{1}{N}(I+\vec{c}.\vec{w})=\frac{1}{N}(I+\Sigma_{j=1}^{2n+1}c_{j}\gamma_{j}),
 \end{equation}
  That is, $\gamma_{j}$ are maximally anticommuting set which satisfy
\begin{equation}
\{\gamma_{i},\gamma_{j}\}=2\delta_{ij}I,
 \end{equation}
  where $I$ stands for the identity operator and $\gamma_{j}$ for $j = 1, 2, . . . , 2n+1,$ known as Dirac matrices, are generators of special orthogonal
group $SO(2n + 1)$, and represented as traceless Hermitian matrices in a $2n$-dimensional Hilbert space.

 Let $N <M$, thus we can represent the density operators acting on a bipartite system $H^{A}\times H^{B
 }$ as:

 \begin{align}
  \rho^{AB}=\frac{1}{NM}(I_{N}\otimes I_{M}+\Sigma_{i,j=1}^{2n+1}t_{ij}\gamma_{i}^{A}\otimes \gamma_{j}^{B}),
 \end{align}
 where, $T=[t_{ij}]$ is the correlation matrix.
 Since quantum correlations are invariant under local unitary transformation, i.e. under
transformations of the form $(U_{1} \otimes U_{2})\rho^{AB}(U_{1}^{\dagger}\otimes U_{2}^{\dagger} )$
with $U_{1},U_{2} \in SU(N)$, we can, without loss of generality,
restrict our considerations to some representative class such that $T$ is diagonal,
namely $T= diag\{t_{1}, t_{2},..., t_{2n+1}\}$ \cite{hor}.
Concerning this representative class, density operators acting on a bipartite system
$H^{A}\times H^{B }$ can be written as:

 \begin{align}\label{ro}
  \rho^{AB}=\frac{1}{NM}(I_{N}\otimes I_{M}+\Sigma_{j=1}^{2n+1}t_{j}\gamma_{j}^{A}\otimes \gamma_{j}^{B}),
 \end{align}

 with eigenvalues
 \begin{align}
  \lambda_{i_{1},i_{2}...i_{2n}}=\frac{1}{NM}[1+(-1)^{i_{1}}t_{1}+(-1)^{i_{2}}t_{2}+...+
  (-1)^{i_{2n}}t_{2n}+(-1)^{n}(-1)^{i_{1}+i_{2}...+i_{2n}}t_{2n+1}],
 \end{align}
 where $i_{1},i_{2},...,i_{2n} \in \{0,1\}$.
 Then the postmeasurement state of $A(B)$ after obtaining the outcome $k$ on $B(A)$, is given by
$$
 \rho_{k}^{A}=\frac{1}{p_{A|k}}Tr_{B}(I_{A}\otimes E_{e^{k}}^{B} \rho^{AB})=\frac{1}{N}(I_{N\times N}+
  \sqrt{\frac{2(M-1)}{M}}\sum_{j}t_{j}e_{j}^{k}\lambda_{j}^{A})
$$
 \begin{align}\label{nn}
  \rho_{k}^{B}=\frac{1}{p_{B|k}}Tr_{A}(E_{e^{k}}^{A}\otimes I_{B}\rho^{AB})=\frac{1}{M}(I_{M\times M}+
  \sqrt{\frac{2(N-1)}{N}}\sum_{j}t_{j}e_{j}^{k}\lambda_{j}^{B}).
 \end{align}

  In the rest of this section we evaluate analytically the quantum discord for any concave quantum entropy functions.
 To begin, consider the following quantum entropy functions \cite{t,s},
 $$
 S(\rho)=-Tr(\rho log_{2}\rho),
 $$

 $$
 S_{R}(\rho)=\frac{1}{1-q}log Tr \rho^{q}, \  (0\leq q\leq1) ,
 $$
\begin{align}\label{q1}
S_{T}(\rho)=\frac{1}{1-q}[Tr\rho^{q}-1 ], \  (0<q) ,
 \end{align}
 respectively the von Neumann, Renyi and Tsallis,  where $S_{R}(\rho)$ and
$S_{T}(\rho)$ are equal to von Neumann entropy in the limit $q = 1$.

 Let us introduce
 \begin{align}
 \mu_{k}=:\sqrt{\frac{2(N-1)}{N}\sum( t_{j}e_{j}^{k})^{2}}.
 \end{align}

 Assume, with no loss of generality , the "left" quantum discord(for the GBSS quantum discord being symmetric between A and B if $N=M$ ).
 Then using the Eqs.(\ref{lef}) and (\ref{q1}) the measure of the classical correlation of the two subsystems $AB$ in the state $\rho^{AB}$ for above quantum entropy functions is given by
\begin{align}
 C_{A}(\rho^{AB})=\max_{\lbrace E_{e^{k}}^{A}\rbrace}\left[logM+\frac{1}{N}\sum_{k} [\frac{1-\mu_{k}}{2}Log(\frac{1-\mu_{k}}{M})+\frac{1+\mu_{k}}{2}
Log(\frac{1+\mu_{k}}{M})]
\right],
\end{align}

$$
 C_{R}(\rho^{AB})= \max_{\lbrace E_{e^{k}}^{A}\rbrace}\left[\frac{1}{1-q}log(M^{-(q-1)})-\frac{1}{M(1-q)}
 \sum_{k} log \frac{M^{-(q-1)}}{2}[(1+\mu_{k})^{q}+(1-\mu_{k})^{q}]\right],
 $$

 \begin{align}
C_{T}(\rho^{AB})=\max_{\lbrace E_{e^{k}}^{A}
\rbrace}\left[\frac{M^{-(q-1)}}{1-q}- \frac{M^{-q}}{2(1-q)}\sum_{k} [(1+\mu_{k})^{q}+(1-\mu_{k})^{q}]\right].
\end{align}

 The above equations represents that the maximum value of the measure of the classical correlation for any kind of entropy only depends on $\mu_{k}$. Thus we turn our attention to the finding maximum value of the $\mu_{k}$.
Hence, the problem of finding the maximum value of the measure of the classical correlation is reduced to the problem

\begin{equation}\label{lp}
\begin{array}{c}
\rm{maximize}\quad \mathcal\cal{\mu_{k}}=\sqrt{\frac{2(N-1)}{N}(\sum t_{j}e_{j}^{k})^{2}},\\
\rm{subject\; to}\quad
e^{k}.e^{k}=1,e^{k}\star e^{k}=e^{k}.
\end{array}
\end{equation}\\

\textbf{Optimization procedure:}\\
Here we present an analytical procedure for
optimization of the measure of the classical correlation for GBSS. This procedure also allows us to obtain the optimum POVMs.

Suppose  $ t_{max}:=max\{|t_{1}|,|t_{2}|,...,|t_{2n+1}|\}$ and its corresponding coefficient in Eq.(\ref{nn}) is $e_{l}^{k}\lambda_{l}^{B}$. Let the set $\{\lambda_{j}^{B}\}$  anticommute  with $\lambda_{l}$ and their related coefficients are $\{e_{j}^{k}\}$ . Choosing these coefficients and making them zero yield a set of optimum POVMs for quantum discord provided that other coefficients chosen such that, the  coefficient  $e_{l}^{k}$ is maximized.
In this case, one can show that the maximum value of $\mu_{k}$ occurs when the all nonzero coefficients are
\begin{align}\label{n}
e_{j}^{k}=\pm\sqrt{\frac{1}{N-1}}.
 \end{align}
 To show this, Let a basis for the Lie algebra of SU(N) be given by

\begin{align}
\{\lambda_{j}\}_{j=0}^{N^{2}-1}=\{\lambda_{\alpha_{1},\alpha_{2},...,
\alpha_{n}}\}
=\{\sqrt{\frac{2}{N}}\sigma_{\alpha_{1}}^{1}\otimes\sigma_{\alpha_{2}}^{2}...
  \otimes\sigma_{\alpha_{n}}^{n}\},
\end{align}
where $\alpha_ {1},\alpha_{2},...,\alpha_{n} \in \{0,1,2,3\}$. For the sake of convenience of analysis, we denote
$$
t_{j}\equiv t_{\alpha_{1}\alpha_{2}...\alpha_{n}},
$$
\begin{align}
e_{j}^{k}\equiv e_{\alpha_{1}\alpha_{2}...\alpha_{n}}^{k},
\end{align}
so Eq.(\ref{lp}) can be written as

 \begin{equation}\label{lp2}
\begin{array}{c}
\rm{maximize}\quad \mathcal\cal{\mu_{k}}=\sqrt{\frac{2(N-1)}{N}(\sum t_{\alpha_{1}\alpha_{2}...\alpha_{n}}e_{\alpha_{1}\alpha_{2}...\alpha_{n}}^{k})^{2}}\\
\rm{subject\; to}\quad
e^{k}.e^{k}=1,e^{k}\star e^{k}=e^{k}.
\end{array}
\end{equation}\\
Using Eq.(\ref{d}) we have
 \begin{align} \label{E}
d_{ijk}=2Tr(\lambda_{i}\lambda_{j}\lambda_{k}),
 \end{align}
 or
 \begin{align} \label{EE}
d_{(\alpha_{1},\alpha_{2},...,
\alpha_{n})(\beta_{1},\beta_{2},...,
\beta_{n})(\gamma_{1},\gamma_{2},...,
\gamma_{n})}=2Tr(\lambda_{\alpha_{1},\alpha_{2},...,
\alpha_{n}}\lambda_{\beta_{1},\beta_{2},...,
\beta_{n}}\lambda_{\gamma_{1},\gamma_{2},...,
\gamma_{n}}),
 \end{align}
 where $\alpha_ {i},\beta_ {i}$ and $\gamma_{i} \in \{0,1,2,3\}$ .

Considering the optimization procedure and using Eqs. (\ref{lp2}) and (\ref{EE}) one can show that the maximum value of $\mu_{k}$ achieve when Eq. $(\ref{n})$ is satisfied.

 Thus, using Eqs. (\ref{n}) and (\ref{lp2}) the maximum value of the $\mu_{k}$ is given by
 \begin{align}\label{nhaa}
\mu_{k,max}=\sqrt{\frac{2}{N}}t_{max}.
 \end{align}
Hence, the measure of the classical correlation due to the von Neumann version entropy is given by

\begin{align}
 C(\rho^{AB})= \frac{1-\mu_{k,max}}{2}Log(1-\mu_{k,max})+
\frac{1+\mu_{k,max}}{2}Log(1+\mu_{k,max})
 \end{align}

Finally, from Eqs. (\ref{d2}), (\ref{d1}) and (\ref{nhaa}), we obtain the quantum discod such as:

\begin{align}\label{nha2}
D(\rho^{AB})=\sum_{j=0}^{N^{2}-1}\lambda_{j} log \lambda_{j}-\frac{1-\mu_{k,max}}{2}Log(1-\mu_{k,max})-
\frac{1+\mu_{k,max}}{2}Log(1+\mu_{k,max})+Log(NM)
 \end{align}
 This is in agreement with the result obtained by Luo in \cite{luo} for $N=M=2$.
  M.D. Lang and C.M. Caves \cite{caves} showed that for the Bell-diagonal states for two qubits
  discord is zero for classical states, which lie on the Cartesian axes and origin.
  While Eq.(\ref{nha2}) shows that for the GBSS, discord is zero only when $t_{1}=t_{2}=...=t_{2n+1}=0$ ($\rho^{AB}=\frac{1}{NM}I$), which lies on the origin. \\
For the Renyi and Tsallis entropy we get
$$
 C_{R}(\rho^{AB})= \frac{1}{1-q}log(M^{-(q-1)})-\frac{1}{1-q}
 log \frac{M^{-(q-1)}}{2}[(1+\mu_{k,max})^{q}+(1-\mu_{k,max})^{q},
$$
 \begin{align}
C_{T}(\rho^{AB})=\frac{M^{-(q-1)}}{1-q}- \frac{M^{-(q-1)}}{2(1-q)}[(1+\mu_{k,max})^{q}+(1-\mu_{k,max})^{q}],
\end{align}
and using Eqs. (\ref{d2}), (\ref{d1}) and (\ref{nhaa}), we obtain the quantum discord such as:
$$
D_{R}(\rho^{AB})=\frac{1}{1-q}log(M^{-(q-1)})-\frac{1}{1-q}log Tr(\rho^{AB})^{q}+ \frac{1}{1-q}
 log \frac{M^{-(q-1)}}{2}[(1+\mu_{k,max})^{q}+(1-\mu_{k,max})^{q}],
 $$
 \begin{align}
D_{T}(\rho^{AB})=\frac{1-M^{-(q-1)}}{q-1}-\frac{1}{1-q}[Tr(\rho^{AB})^{q}]+
\frac{M^{-(q-1)}}{2(1-q)}[-(1+\mu_{k,max})^{q}-(1-\mu_{k,max})^{q}].
\end{align}
Now, we give an exact analytical formula for the POVM. To do this, without loss of generality, we assume that
 $\alpha_ {1},\alpha_{2},...,\alpha_{n} \in \{0,1\}$ and $\beta_ {1},\beta_{2},...,\beta_{n} \in \{0,1\}$.
Then using Eq.(\ref{povm}) the POVM elements are given by

$$
E_{e^{k}}^{A}\equiv E_{\alpha_{1},\alpha_{2},...,\alpha_{n}}=\frac{1}{N}(I-\sqrt{\frac{N}{2}}
\sum_{\beta_{1},\beta_{2},...,\beta_{n}}(-1)^{\sum_{i=1}^{n}
(\alpha_{i}\beta_{i})}\lambda_{\beta_{1},\beta_{2},...,\beta_{n}})
$$
$$
=\frac{1}{N}(I-\sum_{\beta_{1},\beta_{2},...,\beta_{n}}(-1)^{\sum_{i=1}^{n}
(\alpha_{i}\beta_{i})}\sigma_{\beta_{1}}^{1} \otimes \sigma_{\beta_{2}}^{2},...,\otimes \sigma_{\beta_{n}}^{n} )
$$
\begin{align}
=P^{\pm}_{1}\otimes P^{\pm}_{2}\otimes...\otimes P^{\pm}_{n},
\end{align}
where $P^{\pm}_{k}=\frac{1}{2}(I\pm n^{k}.\sigma)$.
 \section{Geometric Measure of Quantum Discord}
 The geometric measure of quantum discord \cite{gmm} is given by:
 \begin{align}
D=\min_{\chi} ||\rho-\chi ||^{2}
\end{align}
 where the minimum is over the set of zero-discord states [i.e., $D(\chi) = 0$]
and the geometric quantity $||\rho-\chi ||^{2}=Tr(\rho-\chi)^{2}$ is the square of Hilbert-Schmidt norm of Hermitian operators.

$\chi$ can be represented as:
\begin{align}
                                 \chi & =\sum_{k}p_{k} |k\rangle\langle |k\otimes\rho_{k},
\end{align}
where $p_{k}$ is a probability distribution, $\{|k\rangle\}$ is an arbitrary orthonormal
basis for $H^{A}$ and $\rho_{k}$ is a set of arbitrary states (density operators) on $H^{B}$ .\\
Consider a bipartite system $H^{A}\otimes H^{B}$ with $dim H^{A} = N$ and $dim H^{B} = M$ . Let $L(H^{A})$
be the space consisting of all linear operators on $H^{A}$. This is a Hilbert
space with the Hilbert-Schmidt inner product
$$
\langle X|Y\rangle=Tr(X^{\dagger} Y)
$$
The Hilbert spaces $L(H^{B})$ and $L(H^{A}\otimes H^{B})$ are defined similarly. Let
$\{X_{i}: i = 1, 2,...,N^{2}\}$ and $\{Y_{j}: j = 1, 2,...,M^{2}\}$ be sets of Hermitian
operators which constitute orthonormal bases for $L(H^{A})$ and $L(H^{B})$
respectively. Then
$$
Tr(X_{i}X_{i^{'}})=\delta_{ii^{'}}, \quad Tr(Y_{j}Y_{j^{'}})=\delta_{jj^{'}}.
$$
$\{X_{i}\otimes Y_{j}\}$ constitutes an orthonormal (product) basis for $L(H^{A}\otimes H^{B})$ (linear
operators on $H^{A}\otimes H^{B}$). In particular, any bipartite state $\rho$ on $H^{A}\otimes H^{B}$
can be expanded as
\begin{align}\label{lug}
    \rho=\sum_{ij}c_{ij}X_{i}\otimes Y_{j} ,
\end{align}
with $c_{ij}=Tr(\rho X_{i}\otimes Y_{j})$.\\
 S. Luo and S. Fu introduced the following form of geometric
measure of quantum discord \cite{lf}
\begin{align}\label{nas}
   D(\rho)=Tr(CC^{t})-\max_{A}Tr(ACC^{t}A^{t}),
\end{align}
where $C=[c_{ij}]$ is an $N^{2}\times M^{2}$ matrix and the maximum is taken
over all $N \times N^{2}$-dimensional isometric matrices $A=[a_{ki}]$ such that
$$
a_{ki}=tr(|k\rangle \langle k|X_{i})=\langle k|X_{i}|k \rangle ,
$$
where $\{|k\rangle\}_{k=1}^{N}$ is any orthonormal base for $H^{A}$.
We can expand the operator $|k \rangle\langle k|$ in the basis of $\{X_{i}\}$ as:
$$
|k \rangle\langle k|=\sum_{i}a_{ki}X_{i}, \quad k=1,2,...,N.
$$
A general bipartite state $\rho$ on $H^{A} \otimes H^{B}$ can be written in this basis as

\begin{align}\label{nasss}
   \rho=\frac{1}{MN}[I_{N}\otimes I_{M}+\vec{x}.\lambda^{A}\otimes I_{M}+I_{N}\otimes \vec{y}.\lambda^{B}+\sum_{i=1}^{N^{2}-1}\sum_{j=1}^{M^{2}-1}t_{ij}\lambda_{i}^{A}\otimes \lambda_{j}^{B}]
\end{align}
where $\vec{x}\in \textbf{R}^{N^{2}-1}$ and $\vec{y}\in \textbf{R}^{M^{2}-1}$ are the coherence
vectors of the subsystems A and B. These are given by
$$
x_{i}=\frac{N}{2}Tr(\rho\lambda_{i}\otimes I_{M}), \quad y_{j}=\frac{M}{2}Tr(\rho I_{N}\otimes\lambda_{j} ).
$$
The correlation matrix $T=[t_{ij}]$ is given by
$$
T=[t_{ij}]=\frac{MN}{4}Tr(\rho\lambda_{i}\otimes\lambda_{j}).
$$
Based on above definitions, Rana et al. \cite{rana} and Hassan et al.\cite{hasan} have shown that the geometric discord of $\rho$ is lower bounded.
By choosing the orthonormal bases $\{X_{i}\}$ and $\{Y_{j}\}$ in Eq.(\ref{lug})
as the generators of $SU(N)$ and $SU(M)$ respectively, that is,
$$
X_{1}= \frac{1}{\sqrt{N}}I_{N}, \quad Y_{1}= \frac{1}{\sqrt{M}}I_{M},
$$
$$
X_{i}=\frac{1}{\sqrt{2}}\lambda_{i-1}, \quad i=2,3,...,N^{2}; \quad\quad Y_{j}=\frac{1}{\sqrt{2}}\lambda_{j-1}, \quad i=2,3,...,M^{2},
$$
 the author\cite{hasan} have shown that
$$
Tr(CC^{t})=\frac{1}{NM}+\frac{2}{M^{2}N}||\vec{y}||^{2}+\frac{2}{N^{2}M}||\vec{x}||^{2}+
\frac{4}{M^{2}N^{2}}||\vec{T}||^{2},
$$
\begin{align}\label{bandttt}
 Tr(ACC^{t}A^{t})=\frac{1}{N}\{\frac{1}{M}+\frac{2}{M^{2}}||\vec{y}||^{2}+\frac{2(N-1)}{N^{2}M}
 [\sum_{j=1}^{N-1}e^{j}G (e^{j})^{t}+\sum_{i=1}^{N-1}\sum_{j=1}^{N-1}e^{i}G (e^{j})^{t}]\},
\end{align}
where $ G=(\vec{x}\vec{x^{t}}+\frac{2TT^{t}}{M} )$ and
$$
e^{k}=\sqrt{\frac{N}{N-1}}(a_{k2},a_{k3},...,a_{kN^{2}}), \quad k=1,2,...,N-1  \quad  \quad ,e^{N}=-\sum_{k=1}^{N-1}e^{k}.
$$

Suppose, for a given pure state $|i\rangle\langle i|$ and an unitary operator U acting on $H^{A}$, there exists an orthogonal operator
$O=[O_{\alpha\beta}]$ acting on $\textbf{R}^{N^{2}-1}$ such that
$$
e^{j}=(n^{j})^{t}O,
$$
where $n^{j}$ is the coherence (column) vector of the state $|i\rangle\langle i|$ and $e^{j}$ is
the coherence (row) vector of the state $U|i\rangle\langle i|U^{\dagger}$.
By writing $G=\sum_{q=1}^{N^{2}-1}\eta_{q}|\hat{f_{q}}\rangle\langle\hat{f_{q}}|$, with its eigenvalues arranged in nondecreasing
order the author\cite{lari} have shown that
\begin{align}\label{lari}
 Tr(ACC^{t}A^{t})=\frac{1}{N}\{\frac{1}{M}+\frac{2}{M^{2}}||\vec{y}||^{2}+\frac{2}{NM}
 \sum_{q=1}^{N^{2}-1}\eta_{q}[\sum_{l=1}^{N-1}|\langle l|O|\hat{f_{q}}\rangle |^{2}]\},
\end{align}
where \{l\} is the standard (computational) basis states in $H^{A}$. The desired maximum is obtained by choosing $O$ in
in Eq.(\ref{lari}) to be that orthogonal matrix which takes the eigenbasis of G matrix to the standard basis in $\textbf{R}^{N^{2}-1}$.
In general, if we consider all of the generators of $SU(N)$ in Eq.(\ref{nasss}) the unitary operator U corresponding to the orthogonal matrix $O$ does not exist. Hence, there exists no an explicit
formula for GMQD in general, but here we show that in the case of the maximally anticommuting set of $\lambda_{j}$ there exists an explicit formula for GMQD.
Consider the maximally anticommuting set of $\lambda_{j}$, that is(for $N<M$)
\begin{align}\label{ro}
  \rho^{AB}=\frac{1}{NM}(I_{N}\otimes I_{M}+\sum_{i=1}^{2n+1}x_{i}\lambda_{i}^{A}\otimes I_{M}+I_{N}\otimes \sum_{i=1}^{2n+1}y_{i}\lambda_{i}^{B}+\sum_{i,j=1}^{2n+1}t_{ij}\gamma_{i}^{A}\otimes \gamma_{j}^{B}).
 \end{align}
 In this case, the unitary operator U exists and it is spinor representation of special orthogonal group $SO(2n+1)$.
 Here, the spinor representations of the group SO(2n + 1) are given by $U=e^{i\sum_{i<j}\theta_{ij}\lambda_{i}\lambda_{j}}$\cite{jafarizadeh2} and
 we can write the transformation of $\gamma_{i}$ more explicitly as
 $$
 U\gamma_{i}U^{\dagger}=\sum_{j=1}^{2n+1}[SO(2n+1)]_{ij}\gamma_{j}.
 $$
Let $a=\sqrt{\frac{N(N-1)}{2}}$ and
\begin{align}\label{band431y}
E_{e^{k}}^{A}=\frac{1}{N}(I_{N\times N}+ae^{k}.\lambda^{A} )=\frac{1}{N}[W_{0}+a(W_{1}+W_{2}+...+W_{N-1})],
\end{align}
where
$$
 W_{0}=I_{N\times N},\quad W_{1}=e'^{k}.\vec{w}=\sum_{i=1}^{2n+1}e'^{k}_{i}\gamma_{i},\quad W_{2}= \sum_{\underbrace{i_{1},i_{2}}_{i_{1}<i_{2}}}
 e_{i_{1}i_{2}}^{k}\gamma_{i_{1}}\gamma_{i_{2}} \quad and
$$
\begin{align}\label{band43y}
W_{N-1}= \sum_{\underbrace{i_{1},i_{2},...,i_{N-1}}_{i_{1}<i_{2}<...<i_{N-1}}} e_{i_{1}i_{2}...i_{N-1}}^{k}\gamma_{i_{1}}\gamma_{i_{2}}...\gamma_{i_{N-1}}.
 \end{align}
 The eigenvectors of $G$ matrix have projection only in the subspace $W_{1}$.
Then, to get the maximum value of $Tr(ACC^{t}A^{t})$, we can choose the orthogonal group $SO(2n+1)$ such that it takes the Bloch vector
components $e^{k}$ in the subspace of maximally anticommuting set $W_{1}$ to the eigenvector of $G$ matrix with the largest eigenvalue.
That is, the orthogonal group $SO(2n+1)$ takes the vector $e'^{k}$ to the eigenvector of $G$ matrix with the largest eigenvalue and
the transformation of the Bloch vector components $e^{k}$ are given by
$$
 e'^{k}_{i_{1}}\rightarrow \sum_{j_{1}} [SO(2n+1)]_{i_{1}j_{1}}e'^{k}_{j_{1}},\quad e_{i_{1}i_{2}}^{k}\rightarrow
 \sum_{j_{1},{j_{2}}}[SO(2n+1)]_{i_{1}j_{1}}SO[(2n+1)]_{i_{2}j_{2}}e_{j_{1}j_{2}}^{k},\quad
$$
$$
e_{i_{1}i_{2}...i_{N-1}}^{k} \rightarrow \sum_{j_{1},...,j_{N-1}}[SO(2n+1)]_{i_{1}j_{1}}...[SO(2n+1)]_{i_{N-1}j_{N-1}}e_{j_{1}j_{2}...j_{N-1}}^{k}.
$$
Then using the optimization procedure of section (\ref{1}) one can show that
\begin{align}\label{band3y}
 \max Tr(ACC^{t}A^{t})=\frac{1}{N}\{\frac{1}{M}+\frac{2}{M^{2}}||\vec{y}||^{2}+\frac{2}{NM}
 \frac{\eta_{max}}{N-1}\},
\end{align}
where $\eta_{max}$ is the largest eigenvalue of the matrix G. Using Eq.s (\ref{nas},\ref{bandttt},\ref{band3y}) we get
\begin{align}\label{band32}
D(\rho)=\frac{2}{N^{2}M}[||\vec{x}||^{2}+\frac{2}{M}||\vec{T}||^{2}-\frac{\eta_{max}}{N-1}].
\end{align}
This is in agreement with the lower bounded obtained in \cite{hasan} for the geometric discord of the bipartite system.

Here, for example , we consider the representation of Gamma matrices in $N = 4$ dimensions. There are different representations for the Gamma matrices, depending on the basis in which they are written. In the Weyl representation (or chiral representation), for any k, we have\cite{jafarizadeh}

 $$
E_{k}^{A}=\frac{1}{N}[I\pm\sum_{i=1}^{5}e_{i}^{k}\gamma_{i}\pm e_{6}^{k}(\sigma_{z}\otimes\sigma_{x})\pm ie_{7}^{k}(\sigma_{z}\otimes\sigma_{y})
\pm ie_{8}^{k}(\sigma_{z}\otimes\sigma_{z})\pm e_{9}^{k}(I\otimes\sigma_{z})\pm
$$
\begin{align}\label{band3111}
e_{10}^{k}(I\otimes\sigma_{y})\pm e_{11}^{k}(I\otimes\sigma_{x})\pm e_{12}^{k}(\sigma_{y}\otimes I)\pm e_{13}^{k}(\sigma_{x}\otimes\sigma_{x})\pm e_{14}^{k}(\sigma_{x}\otimes\sigma_{y})\pm e_{15}^{k}(\sigma_{x}\otimes\sigma_{z})],
\end{align}
where
\begin{align}\label{band366}
\gamma_{1}=\sigma_{x}\otimes I, \quad \gamma_{2}=i\sigma_{y}\otimes \sigma_{x} , \quad \gamma_{3}=i\sigma_{y}\otimes \sigma_{y}, \quad \gamma_{4}=i\sigma_{y}\otimes \sigma_{z}, \quad \gamma_{5}=i\gamma_{1}\gamma_{2}\gamma_{3}\gamma_{4}=-\sigma_{z}\otimes I.
\end{align}
Suppose $e_{5}^{k}\neq 0$, then using the optimization procedure of section (\ref{1}) we have
$$
e^{k}=(0,0,0,0,\pm \frac{1}{\sqrt{3}},0,0,\pm \frac{1}{\sqrt{3}},\pm \frac{1}{\sqrt{3}},0,0,0,0,0,0),
$$
then, we choose the the orthogonal matrix O such that it takes the eigenvector of $G$ matrix with the largest eigenvalue to the  Bloch vector $(0,0,0,0,\pm \frac{1}{\sqrt{3}},0,0,0,0,0,0,0,0,0,0)$.
\section{Geometric interpretation}
Here, we consider $\vec{x}=\vec{y}=0$ and restrict our considerations to some
representative class such that T is diagonal. In this case, using Eq.(\ref{band32}) we get
\begin{align}\label{gbss}
   D(\rho)=\frac{4}{N^{2}M^{2}}[||\vec{T}||^{2}-\frac{t_{max}^{2}}{(N-1)}],
\end{align}
where $||\vec{T}||^{2}=t_{1}^{2}+t_{2}^{2}+...+t_{2n+1}^{2}$.
Now, a complete geometric view is presented for quantum discord and GMQD of GBSS. To do this we determine the feasible region of GBSS . Then we investigate the level surfaces of GMQD.
\subsection{Level surfaces of quantum discord}
M.D. Lang and C.M. Caves \cite{caves} considered the
level surfaces of quantum discord for Bell-diagonal states. They showed that the set of Bell-diagonal states for two qubits can be depicted as a tetrahedron in three dimensions.\\
Here, we point out that the geometric interpretation holds also for generalized Bloch sphere states.
A GBSS are specified by the correlation coefficients $\{t_{1},...,t_{2n+1}\}$. The positivity of $\rho^{AB}$ implies that
\begin{align}
  \lambda_{i_{1},i_{2}...i_{2n}}=\frac{1}{NM}[1+(-1)^{i_{1}}t_{1}+(-1)^{i_{2}}t_{2}+...+
  (-1)^{i_{2n}}t_{2n}+(-1)^{n}(-1)^{i_{1}+i_{2}...+i_{2n}}t_{2n+1}]\geq 0.
   \end{align}
   The above equation gives $N$ inequalities and these inequalities determine the region of GBSS.
   Moreover, by imposing the positivity of partial transposition(PPT) of $\rho^{AB}$, we obtain $N$ other
inequalities. Partial transposition changes the sign of $t_{2n+1}$.
The region defined by intersection of these $2N$ halfspaces is a convex
polytope which is the region of separable GBSS. In fact, the set of separable GBSS are specified by $|t_{1}|+|t_{2}|+...+|t_{2n+1}|\leq 1$.
 On the other hand, the intersection of halfspaces form a convex polytope, where the intersection of its complement and the region of PPT density matrices, is the region of detectable PPT entangled states\cite{jafarizadeh}.

\subsection{Level surfaces of geometric quantum discord}
In order to quantify level surfaces of quantum discord for GBSS ,without loss of generality, we assume that

\begin{align}\label{gbss}
              |t_{1}|>|t_{2}|>...>|t_{2n+1}|,
\end{align}
 that is $t_{max}=t_{1}$, then using Eq.(\ref{gbss}) we have

\begin{align}
              D=\frac{4}{N^{2}M^{2}}[\sum_{i\neq1}t_{i}^{2}+t_{1}^{2}(1-\frac{1}{N-1})].
\end{align}
In the above equation the $D=$constant represents an ellipsoidal region in $2n+1$-dimensional space and Eq.(\ref{gbss}) describes a class of planes.\\
Physical GBSS states belong to the intersection of these planes and ellipsoid.\\
In fact the level surfaces of geometric discord of GBSS are composed of $2n+1$
identical intersecting ellipsoids. For the set of Bell-diagonal states for two qubits the level surfaces of geometric discord are composed of three identical intersecting cylinders instead of ellipsoids\cite{zzz}.
When the D decreases ellipsoids shrink towards the origin and when the D increases the ellipsoids increase outward from the origin such that in the region of GBSS the maximum value of D occurs when $|t_{1}|=...=|t_{2n+1}|=1$. So we have
\begin{align}
              D_{max}=\frac{4}{N^{2}M^{2}}(2n+1-\frac{1}{N-1}).
\end{align}

The separable states is actually bounded by  $|t_{1}|+|t_{2}|+...+|t_{2n+1}|\leq 1$.
 Since in the region of separable GBSS $||T||^{2}\leq 1$, then the maximum value of the $D$ occurs, when $||T||^{2}=1$ and $t_{max}$ is as small as possible. This leads
$$
|t_{1}|=|t_{2}|=...=|t_{2n+1}|=\frac{1}{\sqrt{2n+1}},
$$
and the maximum value of the geometric measure of quantum discord for GBSS is given by
\begin{align}
              D_{max}=\frac{4}{N^{2}M^{2}}(1-
             \frac{1}{(2n+1)(N-1)}).
\end{align}
 This is in agreement with the result obtained in \cite{gmm} for the set of Bell-diagonal states for two qubits.\\

\section{Conclusion}
In summary, we have developed a complete and intuitive analytic picture of the quantum discord problem
 for GBSS that lends itself to a straightforward analytic
 algorithm to finding the optimal POVM. In addition, we have shown that the result does not depend on the
 entropy function. That is the same POVM is optimum for measuring quantum discord of GBSS for any concave entropy function in the definition of quantum discord.
 Also, we have pointed out that the presented procedure of optimization is also used to find the geometric measure of quantum discord and we provide an analytical expression for the geometric quantum discord.
 Finally, we have developed the geometric interpretation of original quantum discord and geometric measure of quantum discord for GBSS.
\vspace{1cm} \setcounter{section}{0}
 \setcounter{equation}{0}
 \renewcommand{\theequation}{I-\arabic{equation}}
\newpage
{\Large{Appendix :}}\\

Throughout the paper, we have used the formalism of Dirac $\gamma$ matrices. Therefore, in this appendix we define the algebra of Dirac $\gamma$ matrices and exhibit matrices which
realize the algebra in the Euclidean representation and explain our notations and conventions.\\
To do this, let $\gamma_{\mu}, \mu = 1, ..., d,$ be a set of d matrices satisfying the anticommuting relations:
\begin{align}\label{e}
\gamma_{\mu}\gamma_{\nu}+\gamma_{\nu}\gamma_{\mu}=2\delta_{\mu\nu}I,
\end{align}
in which I is the identity matrix.
These matrices are the generatores of a Clifford algebra similar to the algebra of operators
acting on Grassmann algebras. It follows from relations (\ref{e}) that the $\gamma$ matrices generate an
algebra which, as a vector space, has a dimension $2^{d}$. In the following, we will give an inductive
construction $(d \rightarrow d+2)$ of hermitian matrices satisfying (\ref{e}). In the algebra one element
plays a special role, the product of all $\gamma$ matrices. The matrix $\gamma_{s}$:
\begin{align}
\gamma_{s}=i^{\frac{-d}{2}}\gamma_{1}\gamma_{2}...\gamma_{d},
\end{align}
anticommutes, because $d$ is even, with all other $\gamma$ matrices and $\gamma_{s}^{2}= I$.

In calculations involving $\gamma$ matrices, it is not always necessary to distinguish $\gamma_{s}$ from other
$\gamma$ matrices. Identifying thus $\gamma_{s}$ with $\gamma_{d+1}$, we have:
\begin{align}\label{ee}
\gamma_{i}\gamma_{j}+\gamma_{j}\gamma_{i}=2\delta_{ij}I, i,j=1,...,d,d+1.
\end{align}
The Greek letters $\mu\nu...$ are usually used to indicate that the value $d+1$ for the index has been
excluded.

\textbf{An explicit construction of $\gamma_{i}^{(d)}$}\\

It is sometimes useful to have an explicit realization of the algebra of $\gamma$ matrices.
For $d = 2$, the standard Pauli matrices realize the algebra:
$$
\gamma_{1}^{(d=2)}=\sigma_{1}=\left(
             \begin{array}{cc}
               0 & 1 \\
               1 & 0 \\
             \end{array}
           \right) , \gamma_{2}^{(d=2)}=\sigma_{2}=\left(
             \begin{array}{cc}
               0 & -i \\
               i & 0 \\
             \end{array}
           \right),
$$
\begin{align}
\gamma_{s}^{(d=2)}= \gamma_{3}^{(d=2)}=\sigma_{3}=\left(
             \begin{array}{cc}
               1 & 0 \\
               0 & -1 \\
             \end{array}
           \right)
\end{align}
The three matrices are hermitian, i.e., $\gamma_{i}=\gamma_{i}^{\dagger}$. The matrices $\gamma_{1}$ and $\gamma_{3}$ are symmetric and $\gamma_{2}$ is antisymmetric, i.e., $\gamma_{1}=\gamma_{1}^{t}$,  $\gamma_{3}=\gamma_{3}^{t}$ and $\gamma_{2}=-\gamma_{2}^{t}$.
To construct the ã matrices for higher even dimensions, we then proceed by induction,
setting:
$$
\gamma_{i}^{(d+2)}=\sigma_{1}\otimes \gamma_{i}^{(d)}=\left(
                                                          \begin{array}{cc}
                                                            0 & \gamma_{i}^{(d)} \\
                                                            \gamma_{i}^{(d)} & 0 \\
                                                          \end{array}
                                                        \right), i=1,2,...,d+1,
$$
\begin{align}\label{eee}
\gamma_{d+2}=\sigma_{2}\otimes I^{(d)}=\left(
                                          \begin{array}{cc}
                                            0 & -iI_{d} \\
                                            iI_{d} & 0 \\
                                          \end{array}
                                        \right),
                                        \end{align}
where, $I_{d}$ is the unit matrix in $2^{\frac{d}{2}}$ dimensions.
As a consequence $\gamma_{s}^{(d+2)}$ has the form:
\begin{align}
\gamma_{s}^{(d+2)}=\gamma_{d+3}^{(d+2)}=\sigma_{3}\otimes I_{d}=\left(
                                                                    \begin{array}{cc}
                                                                      I_{d} & 0 \\
                                                                      0 & -I_{d} \\
                                                                    \end{array}
                                                                  \right)
\end{align}
A straightforward calculation shows that if the matrices $\gamma_{i}^{(d)}$ satisfy relations (\ref{ee}), the $\gamma_{i}^{(d+2)}$ matrices satisfy the same relations. By induction we see that the $\gamma$ matrices are all hermitian. from (\ref{eee}), it is seen that, if $\gamma_{i}^{(d)}$ is symmetric or antisymmetric, $\gamma_{i}^{(d+2)}$
 has the same property.
 The matrix $\gamma_{d+2}^{(d+2)}$ is antisymmetric and $\gamma_{s}^{(d+2)}$ S which is also $\gamma_{d+3}^{d+2}$
 is symmetric. It follows immediately that, in this representation, all $\gamma$ matrices with odd index are symmetric and all matrices with even index are antisymmetric, i.e.,
\begin{align}
\gamma_{i}^{t}=(-1)^{i+1}\gamma_{i}.
\end{align}

\end{document}